\documentclass[10pt,journal]{IEEEtran}

\usepackage[latin1]{inputenc}
\usepackage{cite}
\usepackage{amsmath}
\usepackage{subfigure}
\usepackage[figure,linesnumbered]{algorithm2e}

\usepackage{reledmac}

\usepackage[per-mode=symbol]{siunitx}
\usepackage{ifpdf}
\ifpdf
  \usepackage[pdftex]{graphicx}
  \else \usepackage[dvips]{graphicx}
\fi
\usepackage{booktabs}
\usepackage[flushleft, para]{threeparttable}

\usepackage{xcolor}

\newcommand\BeraMono{%
  \def\fvm@Scale{0.80}% scales the font down
  \fontfamily{fvm}\selectfont% selects the Bera Mono font
}

\title{Hardware Implementation of an OPC UA Server for Industrial Field Devices}

\author{Heiner Bauer, Sebastian Höppner, Chris Iatrou, Zohra Charania, Stephan Hartmann, Saif-Ur Rehman, Andreas Dixius, Georg Ellguth, Dennis Walter, Johannes Uhlig, Felix Neumärker, Marc Berthel, Marco Stolba, Florian Kelber, Leon Urbas, Christian Mayr
\thanks{Manuscript received \today. This work was supported by the BMBF project fast semantics (FKZ: 03ZZ0521D).
}%
\thanks{The authors are with the Faculty of Electrical and Computer Engineering, Technische Universität Dresden, Germany (e-mail: heiner.bauer@tu-dresden.de)}
}

\usepackage[hidelinks]{hyperref}

\begin{document}
\bstctlcite{IEEEexample:BSTcontrol}

\maketitle

% 100 - 250 words
\begin{abstract}
% IEEE TVLSI: 100 to max. 250 words
Industrial plants suffer from a high degree of complexity and incompatibility in their communication infrastructure, caused by a wild mix of proprietary technologies.
This prevents transformation towards Industry 4.0 and the Industrial Internet of Things.
Open Platform Communications Unified Architecture (OPC UA) is a standardized protocol that addresses these problems with uniform and semantic communication across all levels of the hierarchy.
However, its adoption in embedded field devices, such as sensors and actors, is still lacking due to prohibitive memory and power requirements of software implementations.
We have developed a dedicated hardware engine that offloads processing of the OPC UA protocol and enables realization of compact and low-power field devices with OPC UA support.
As part of a proof-of-concept embedded system we have implemented this engine in a 22 nm FDSOI technology.
We measured performance, power consumption, and memory footprint of our test chip and compared it with a software implementation based on open62541 and a Raspberry Pi 2B.
Our OPC UA hardware engine is 50 times more energy efficient and only requires 36 KiB of memory. The complete chip consumes only 24 mW under full load, making it suitable for low-power embedded applications.

\end{abstract}

% at least 5 keywords

\setlength{\textfloatsep}{10pt}

\section{Introduction}
\begin{figure*}
  \centering
  \includegraphics[width=0.9\textwidth]{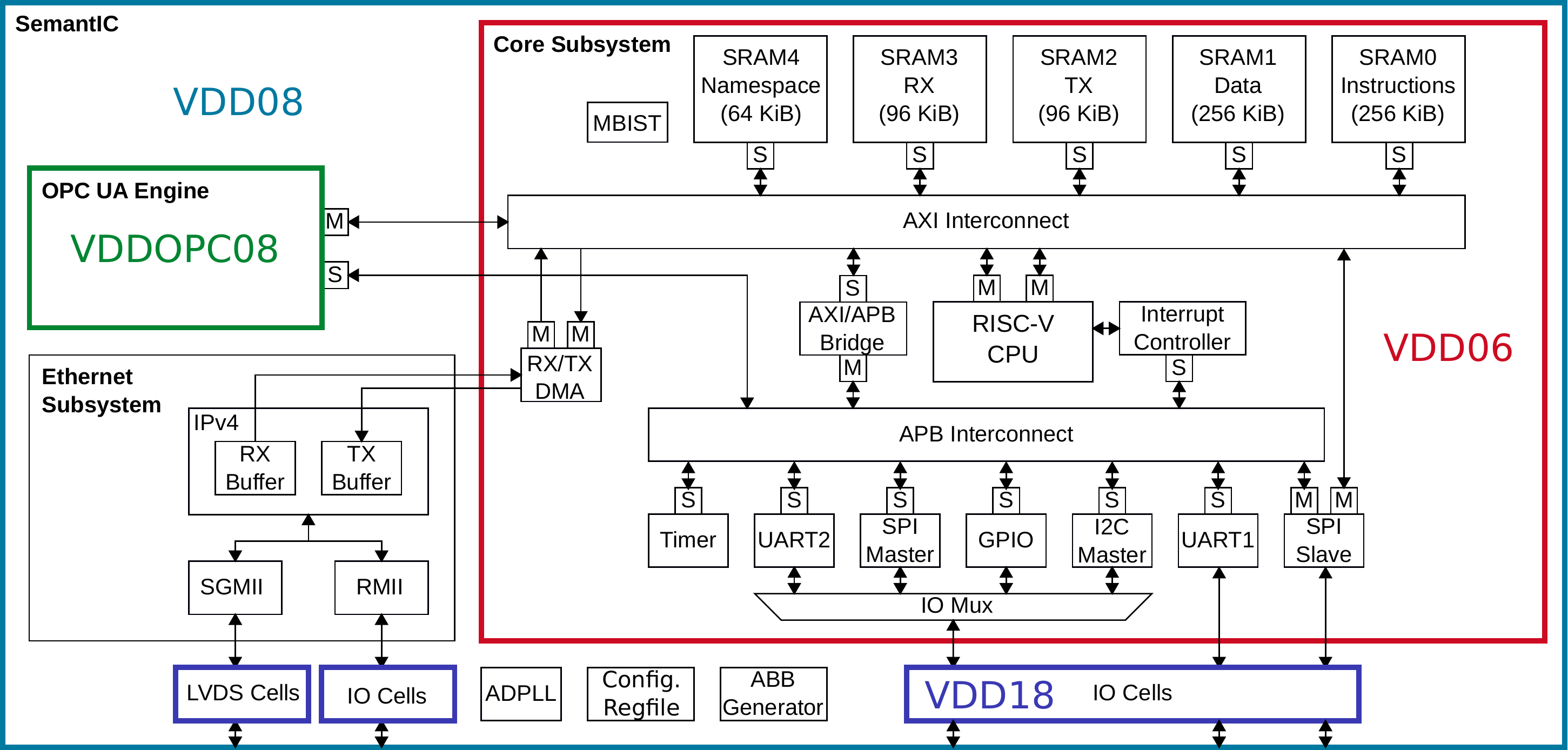}
  \caption{Simplified block diagram of the test chip. Colors mark the power domains that can be measured separately on package level. SRAM in the VDD06 domain uses a separate rail for the bitcells that is connected to VDD08.}
  \label{fig:system-diagram}
\end{figure*}
The communication infrastructure in most industrial plants consists of multiple hierarchical layers and is highly segregated. Especially the lower layers with control systems and field devices employ a wild mix of largely proprietary communication technologies \cite{bib:Urbas2012}. Examples include analog current loops, serial protocols such as HART, IOLink, or Modbus RTU, but also technologies based on Ethernet like ProfiNet or Modbus TCP. Apart from the difficulty to maintain such complex and heterogeneous systems, the compartmentalization and incompatibilities between these technologies severely restrict the flow of information. This leads to problems with current trends such as Industry 4.0 and Industrial Internet of Things \cite{bib:Atzori2010} which demand safe and secure access to information and meta-information from all layers to further optimize performance, efficiency, and yield of industrial processes.

OPC UA (Open Platform Communications Unified Architecture) is a platform-independent communication standard \cite{bib:IEC-OPCUA} with extensive capabilities for semantic information modeling that directly addresses these problems.
It functions as a middleware on top of universal and established transport protocols such as Ethernet and uses a client-server model with standardized services such as read-write access to data, discovery of servers and their capabilities, subscription to value changes, authentication, and encryption.
The data exchange between devices is enriched with semantic representations by the OPC UA information model, which describes the attributes and relationships of objects on the server.
For example, an OPC UA enabled pressure sensor could convey the measurement unit, measurement range, sensor type, serial number, manufacturer, and even measurement quality metrics in addition to just the raw pressure value. Instead of an arbitrary list of values, this information is structured with an object oriented view using typed references that describe the properties of the equipment.

The OPC UA information model is based on a generic meta-model, called namespace 0, that provides a powerful and common vocabulary for semantic descriptions \cite{bib:Graube2017}.
Vendor-independent interoperability is achieved by standardizing this meta-model as well as domain-specific extensions and by embedding the model into the devices.

OPC UA has been widely adopted and all major vendors of industrial equipment offer products that support OPC UA to various degrees.
But, up until now, OPC UA has only penetrated the upper layers in the communication hierarchy and few examples of embedded OPC UA field devices exist, e.g. \cite{bib:Imtiaz2013}. We believe this is largely due to the complexity and resource requirements imposed by software implementations of the protocol.

These software implementations often require a complete operating system with a networking stack and virtual memory management, and thus can have a significant memory (several hundred kilobytes) and power (several hundred milliwatts) footprint.
Powerful application-class SoCs with memory controllers and external DRAM are usually needed to support these software implementations.
This can incur significant overhead in terms of cost, PCB area, and energy consumption, which prevents adoption of OPC UA in space or energy constrained applications.
For instance, field devices in explosion hazardous environments must be as power efficient as possible to avoid hot surfaces and are heavily space constrained by costly enclosures.
Additionally, most OPC UA software implementations are unsuitable for hard real-time applications due to jitter caused by interactions of the operating system and applications running on the shared hardware.

We present a dedicated hardware engine that implements an OPC UA server as part of a compact and low-power embedded system.
Because processing of the OPC UA protocol is completely offloaded, this approach also has the potential to satisfy hard real-time constraints.
To evaluate our OPC UA hardware solution and to provide a proof-of-concept system for OPC UA enabled field devices, we have implemented this engine on a \SI{22}{\nano\metre} test chip. To the best of our knowledge, this constitutes the first ever ASIC implementation of an OPC UA server for field devices.

\section{System Architecture}
The block diagram in figure \ref{fig:system-diagram} gives an overview of the test chip architecture. The chip features all major components required for an industrial field device with OPC UA capabilities. It is derived from the concept that has been described in \cite{bib:Iatrou2019}.
\subsection{Core Subsystem}
The main component in the core subsystem is a 32-bit RISC-V CPU with support for IMCXpulpv2 instructions \cite{bib:Gautschi2017}. Five independent SRAM banks are attached to an AXI interconnect structure that assigns priority to the CPU for SRAM0 and SRAM1 for storing application code and data, priority for the Ethernet DMA to SRAM2 and SRAM3, and priority to the OPC UA engine for SRAM4 to store the information model of the field device.
Standard serial peripherals such as UART, SPI, and I2C are attached to the CPU so that applications can interact with external sensor and actor devices.
% GPIO mux with debug mode and software controlled GPIOs ?
A dedicated SPI slave is used to configure the chip after power up and initialize the volatile main memory from an external source.
\subsection{Ethernet Subsystem}
The Ethernet subsystem implements a tri-mode MAC, supporting 10, 100, and 1000 MBit/s, as well as packet processing for IPv4, ICMP, and UDP. A DMA transfers data between buffers in the IPv4 block and the main SRAM, without requiring the CPU to act on every sent or received packet.
An SGMI interface with an LVDS SerDes provides a connection to external gigabit PHYs. The SerDes is clocked from an all-digital PLL on top level that has 8 phase outputs and an internal DCO running at \SI{2.5}{\giga\hertz}.
We also included an RMI interface to evaluate emerging Ethernet solutions which are tailored to long-reach industrial field communications, e.g. Advanced Physical Layer based on 10BASE-T1L.

% short but plausible description how the OPC UA protocol is processed in hardware
\subsection{OPC UA Engine}
\begin{figure}
  \centering
  \includegraphics[width=0.4\textwidth]{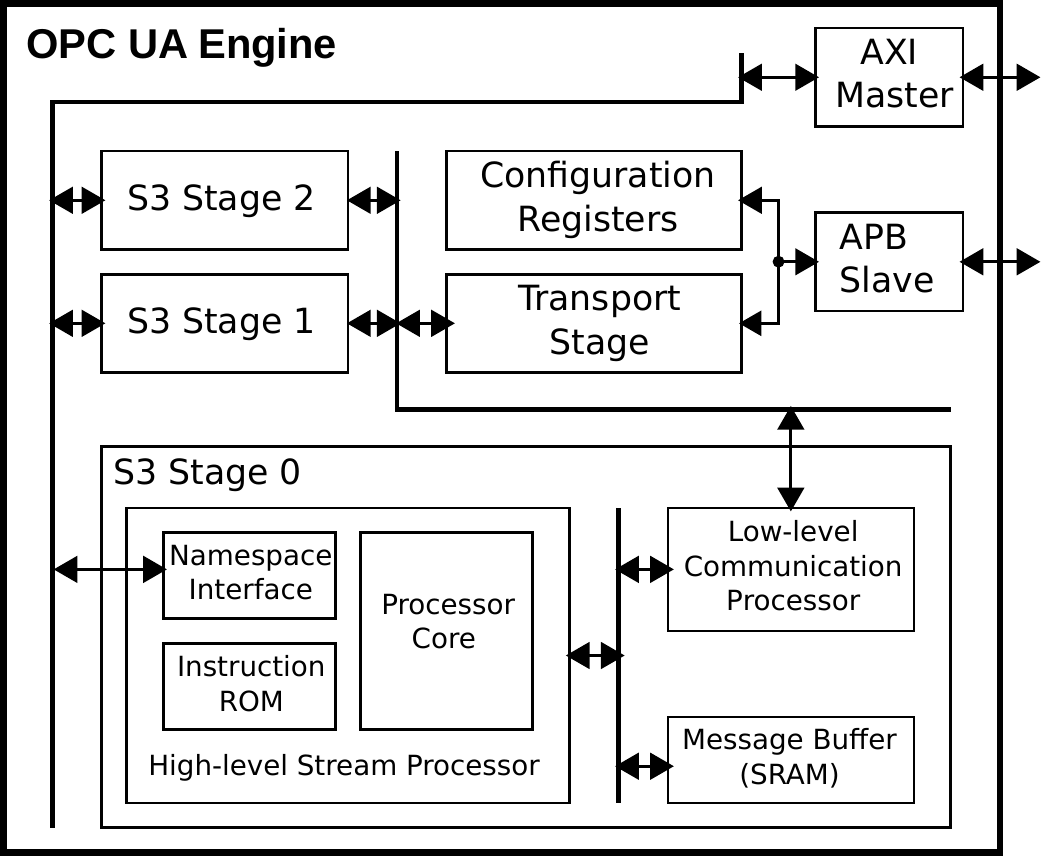}
  \caption{Internal architecture of the OPC UA Engine with details shown for one S3 stage.}
  \label{fig:opcua-diagram}
\end{figure}
Details of the internal architecture of the OPC UA engine are shown in Figure \ref{fig:opcua-diagram}. The engine implements the functionality of an OPC UA server  with the Nano Embedded Device profile, but offers support for three sessions in the shown configuration. The design offers extensive design-time configuration options, for example, the size of the internal SRAM, message fragmentation support, the number of parallel sessions, and the supported OPC UA session features.

OPC UA requests are sent and received by a transport stage that is attached to the peripheral bus of the system. This bus interface also allows to the CPU to access internal registers that control operation of the engine.
The transport stage directly handles basic operations to establish sessions and to advertise the supported protocol versions, buffer sizes, and transport layer capabilities.

More complex requests are forwarded to one of the available S3 (security, segmentation, and services) stages.
Each S3 stage can maintain and serve one distinct OPC UA sessions. S3 stages are compromised of three major components that are interconnected with a multi-mastered address bus.
A low-level communication processor handles message reception and transmission, with emphasis on message segmentation or ``chunking''.
The second major component is the SRAM message buffer where message assembly takes place.

OPC UA service requests (e.g. read node) are encoded akin to serialized object-oriented data structures, implying that message fields present for any single service request type may differ radically depending on parameters.
The third component, a specialized, high-level OPC UA stream processor, interprets these service requests and formulates the reply message.
The processor uses a Harvard architecture and natively handles OPC UA's numerous data types and their highly irregular encoding.
This includes multi-part types (e.g. the "LocalizedText" type, composed of two strings), variable encoding types (e.g. "NodeID"), and even user-defined data types ("Structures").
To save space, the instruction set uses variable-length encoding.
The processor supports copy operations of all OPC UA data types between up to 16 streams, most prominently the address space and the message.
Logical comparison between OPC UA data types is supported along with simple branching and call/return mechanics together with a hardware stack-pointer.
The instruction set supports creating indexes on stream positions, allowing a service program to seek specific positions in a data stream.
The development of service programs was simplified with a custom assembler written in Python, which translates the instruction sequences into synthesizable Verilog code. This is represented in figure \ref{fig:opcua-diagram} as the instruction ROM.

The namespace interface allows service programs to access an object or node in the OPC UA information model.
It understands the structure of the namespace image in the main memory, which uses a custom binary encoding to reduce its memory footprint \cite{bib:Iatrou2016a}.
The namespace interface autonomously performs the node lookup in the main memory (SRAM4) and returns the requested data to the stream processor.

Communication between the transport stage, the S3 stages, and the main memory is scheduled with a fair and deterministic scheme that provides the basis for hard real-time guarantees of the OPC UA engine.
Further details regarding the processing structure of the OPC UA engine are discussed in a previous publication \cite{bib:Iatrou2016b}.
%%% Local Variables:
%%% mode: latex
%%% TeX-master: "../paper_semantic"
%%% End:
\section{Test Chip Implementation}
\begin{figure}
  \centering
  \includegraphics[width=0.35\textwidth]{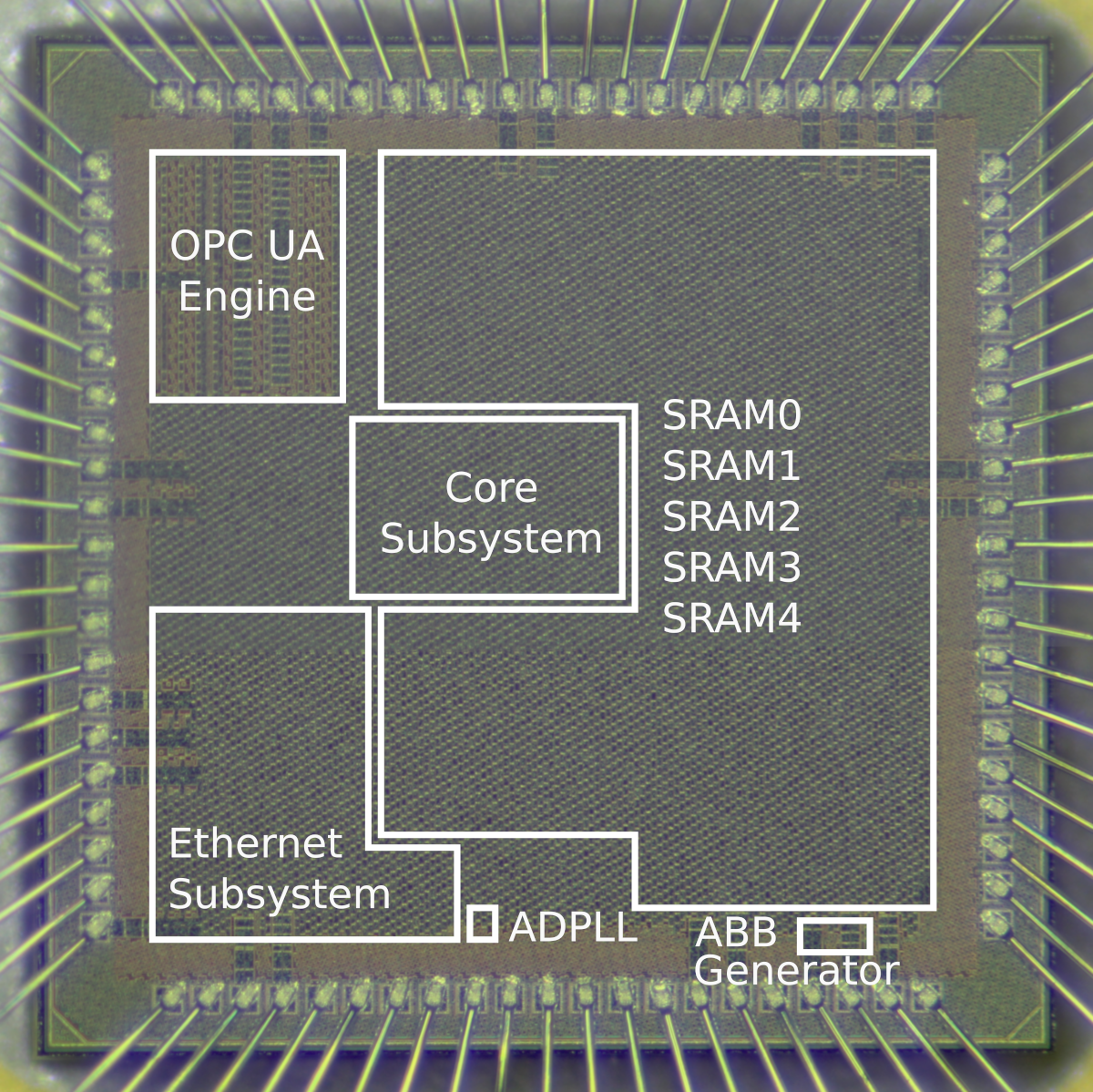}
  \caption{Annotated photo of the SemantIC test chip.}
  \label{fig:chip-photo}
\end{figure}

Figure \ref{fig:chip-photo} shows a photo of the test chip, which has been fabricated in a GLOBALFOUNDRIES 22 nm FDSOI technology \cite{bib:Carter2016}.
We leverage the body biasing feature of the FDSOI technology with a closed-loop, on-chip adaptive body bias (ABB) generator.
Together with appropriately characterized standard cell and SRAM libraries, this guarantees robust performance at minimal leakage over all PVT corners \cite{bib:Hoeppner2020}.
Implementation and signoff of the test chip has been carried out for the full industrial temperature range, from \SI{-40}{\celsius} to \SI{125}{\celsius}.

The ABB generator applies a forward bias to the core subsystem and enables us to implement the RISC-V CPU with \SI{250}{\mega\hertz} at only \SI{0.6}{\volt}. This reduced supply voltage necessitates dual-rail SRAM, where the bitcells are supplied from the top level with \SI{0.8}{\volt}. Peripherals operate from the same \SI{0.6}{\volt} supply but at a reduced clock frequency of 100 MHz.
The rest of the chip, including the OPC UA engine, operates without adaptive body bias.

\begin{figure}
  \includegraphics[width=0.5\textwidth]{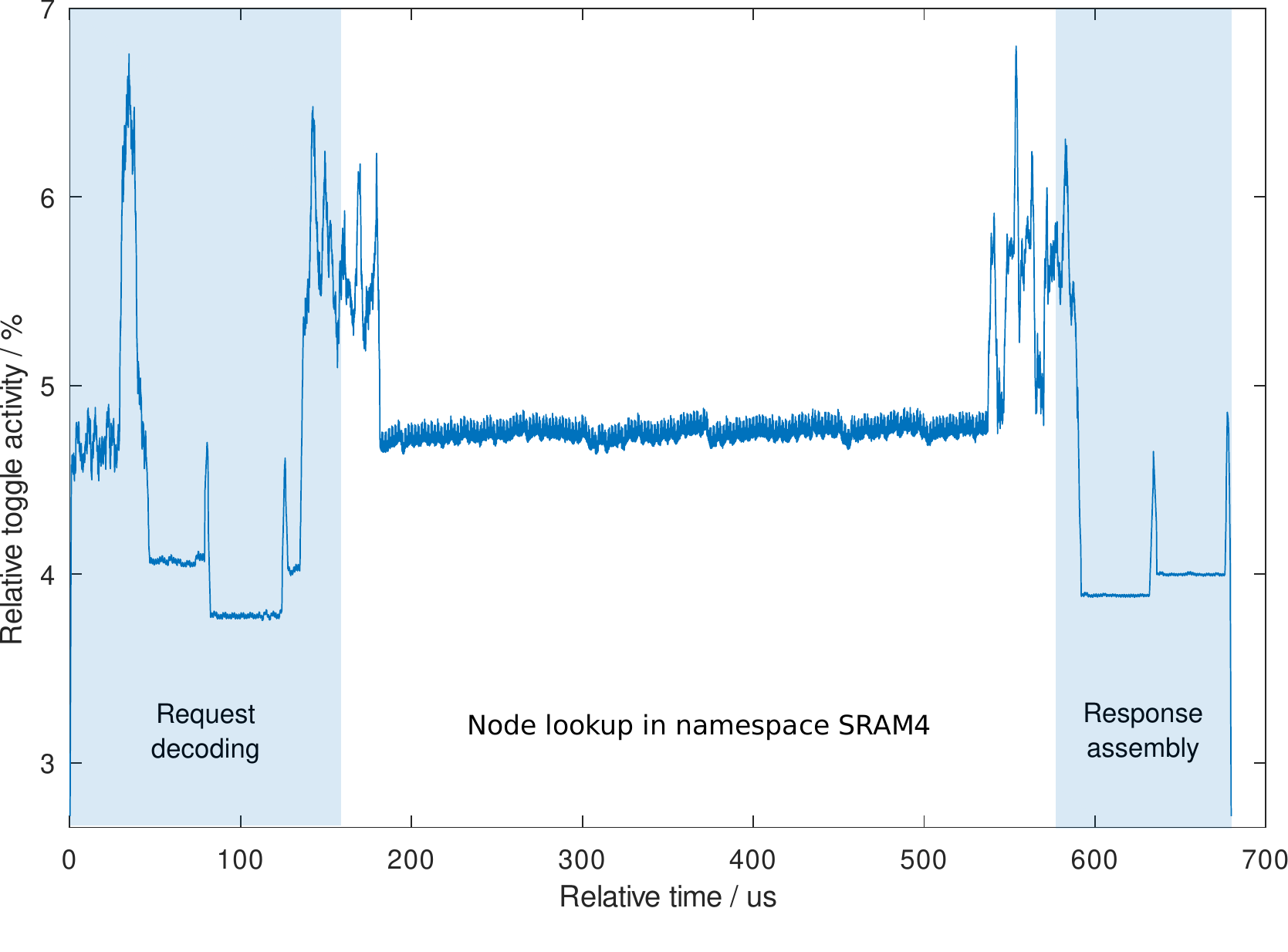}
  \caption{Activity profile of the OPC UA engine during a read node operation captured from a netlist simulation. The average activity is \SI{4.7}{\percent}.}
  \label{fig:acitivity-profile}
\end{figure}

Because the OPC UA engine processes messages serially, this block has a low toggle activity.
Figure \ref{fig:acitivity-profile} shows an activity profile of the OPC UA engine during processing of a read node message (the write node activity profile is nearly identical). The average activity during this window is only \SI{4.7}{\percent}.
Additionally, the engine can be completely clock-gated while waiting for new OPC UA messages to arrive.
Therefore our goal was to minimize leakage and we restricted synthesis of the OPC UA engine to use only HVT cells. With this we were able to close timing in the final implementation at \SI{50}{\mega\hertz} with over \SI{98}{\percent} HVT cells.

The complete macro of the OPC UA engine with three S3 stages occupies \SI{200000}{\square\micro\metre}. This includes 24 KiB internal SRAM
for the message buffers and 269 kGE (kilo gate equivalents) for the remaining logic, with 78 kGE per S3 stage.

We used the open-source ICGlue\footnote{\href{https://www.icglue.org}{www.icglue.org}} generator to construct the hardware description of the top level chip hierarchy and the configuration register files that control the ADPLL, ABB generator, and IO cells. ICGlue also generates documentation and software abstractions for these registers which accelerates chip design and verification, but also bring up and measurement in the lab.

\section{Measurement Results}
We measured power consumption of four chip samples at room temperature with an Agilent B2962A power supply, which supports high precision current and voltage sensing.

All measurements related to the OPC UA engine were performed with the minimal namespace 0 from open62541\footnote{\href{https://www.open62541.org}{www.open62541.org}}, a popular OPC UA software implementation, and a custom device information model with three 32-bit values. Using our optimized binary format this model occupies only 36 KiB in SRAM4.
For our scenario of embedded sensor and actor field devices, read and write requests represent the most relevant OPC UA services. Therefore, we stimulated the OPC UA engine by running an application on the CPU that uses prerecorded OPC UA messages to establish a session and then continuously reads or writes the last node in our device model.
\begin{figure}
  \includegraphics[width=0.5\textwidth]{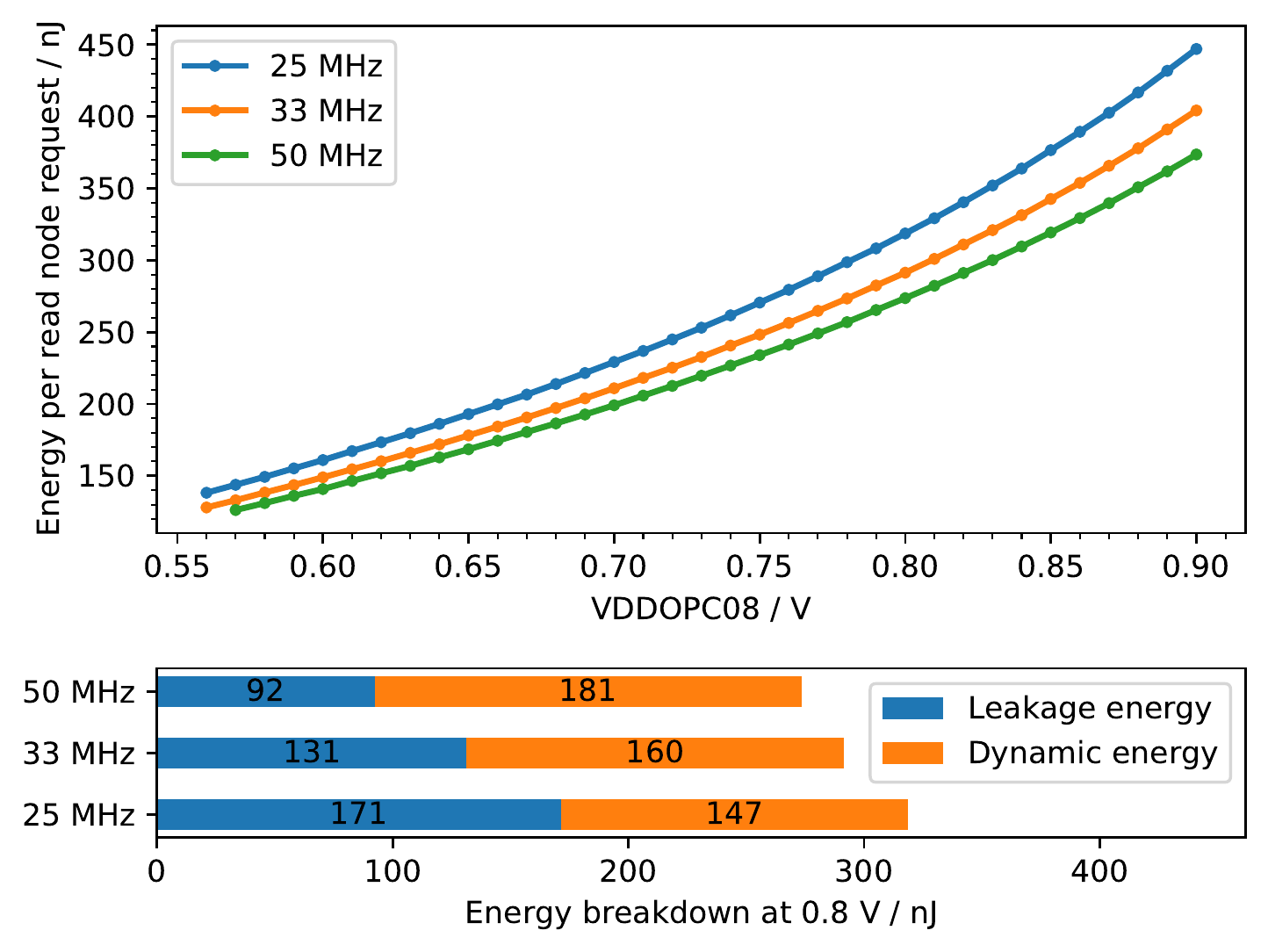}
  \caption{Energy consumption of the OPC UA engine for a read node request.}
  \label{fig:opcua-vdd-sweep}
\end{figure}
We measured the power consumption during the read node requests for three different operating frequencies and over a voltage sweep of the VDDOPC08 domain and calculated the energy for each point as shown in figure \ref{fig:opcua-vdd-sweep}. The engine processes a read node (write node) request within \SI{1219}{\micro\second} (\SI{1186}{\micro\second}) at \SI{25}{\mega\hertz}, \SI{935}{\micro\second} (\SI{910}{\micro\second}) at \SI{33}{\mega\hertz}, and \SI{657}{\micro\second} (\SI{638}{\micro\second}) at \SI{50}{\mega\hertz}.
Note that these latencies don't scale exactly with the frequency of the OPC UA engine due to the access synchronization with the main memory, which runs at a fixed frequency.
With the nominal supply voltage of \SI{0.8}{\volt} the engine requires \SI{273}{\nano\joule} to process a read node request and \SI{266}{\nano\joule} to process a write node request.

In table \ref{tab:power-results-hardware} we give further measurement data to characterize the power consumption of the complete system.
We ran the CoreMark benchmark on the CPU to evaluate the energy efficiency of the core subsystem.
To stimulate network load and measure the power consumption of the Ethernet subsystem, we used an external PC and ran a flood ping to the test chip in SGMII mode.

\setlength{\tabcolsep}{0.3em} % for the horizontal padding
\begin{table}[h]
  \centering
  \caption{Test Chip Power Consumption}
  \begin{threeparttable}
    \begin{tabular}{@{}llrrrrr@{}}
      \toprule
           & Scenario         & VDDOPC08 & VDD08 & VDD06 & VDD18 & Units           \\
      \midrule
      0     & Leakage          & 121      & 390   & 342   & 144   & \si{\micro\watt}\\
      1     & Idle, clocks on  & 0.418    & 3.23  & 1.08  & 15.98 & \si{\milli\watt}\\
      \midrule
      2      & CoreMark     &     -    & 3.87  & 2.30  &  -    & \si{\milli\watt}\\
      3      & Read node    & 0.426    & 3.28  &     - &  -    & \si{\milli\watt}\\
      4      & Write node   & 0.427    & 3.28  &     - &  -    & \si{\milli\watt}\\
      5      & Flood ping   &     -    & 4.18  &     - & 16.44 & \si{\milli\watt}\\
      \bottomrule
    \end{tabular}
  \end{threeparttable}
  \label{tab:power-results-hardware}
\end{table}

The total power consumption of the test chip $P_{total}$ can be calculated by summing the OPCUA read power $P_{3, VDDOPC08}$, the power in the core subsystem $P_{2, VDD06}$, the power of toplevel components $P_{5, VDD08}$, and the IO cells $P_{5, VDD18}$.  Additionally, we need to account the differential power of OPC UA activity in SRAM4 with $P_{3, VDD08} - P_{1, VDD08}$, and CPU activity in SRAM0 and SRAM1 with $P_{2, VDD08} - P_{1, VDD08}$.
With this, $P_{total} = \SI{24.04}{\milli\watt}$ represents the active power of the complete system under load. Only 1.8 \% of this are due to the OPC UA engine.

For a comparison we could not find published data for OPC UA software implementations concerning power and energy consumption in an embedded field device.
Thus we decided to run our own reference measurements with the open62541 software stack and a Raspberry Pi 2B. % Model B v1.2
The Raspberry Pi platform is widely available, easy to set up with open62541, and also used in several embedded industrial products.
This particular board uses a quad-core Arm Cortex-A53 SoC (BCM2837), a 1 GiB LPDDR2 DRAM (EDB8132B4PB), and a USB hub with integrated Ethernet PHY (LAN9514).
We configured Linux to run with a fixed CPU frequency and enabled only the bare minimum of processes required to run open62541.
Next, we compiled open62541 in version v1.0.5 with the recommended options to minimize memory consumption.
\begin{table}[h]
  \centering
  \caption{Raspberry Pi 2B Power Consumption}
  \begin{threeparttable}
    \begin{tabular}{@{}lrrr@{}}
      \toprule
           Scenario              & 600 MHz & 900 MHz & Units           \\
      \midrule
      SoC idle                   & 1.186   & 1.198  & \si{\watt}\\
      SoC idle, LAN9514 disabled & 0.392   & 0.420  & \si{\watt}\\
      SoC idle, network load     & 1.231   & 1.247  & \si{\watt}\\
      OPC UA read node           & 1.269   & 1.316  & \si{\watt}\\
      OPC UA write node          & 1.271   & 1.327  & \si{\watt}\\
      \midrule
      Estimated SoC read node power\tnote{1}     & 18.9        &  34.2      & \si{\milli\watt}\\
      Estimated SoC write node power\tnote{1}    & 20.0        &  40.1      & \si{\milli\watt}\\
      \bottomrule
    \end{tabular}
    \begin{tablenotes}
      \item[1] Assuming worst-case 50 \% efficiency of the DC-DC converters
    \end{tablenotes}
  \end{threeparttable}
  \label{tab:power-results-software}
\end{table}
The \texttt{top} utility reported 480 KiB of resident memory occupied by open62541 while handling one OPC UA session.
Such a large memory footprint would be prohibitive for most embedded SoCs without external DRAM.

We present the power consumption of the Raspberry Pi, measured at the 5 V input of the USB connector, in the first half of table \ref{tab:power-results-software}.
To account for losses in the DC-DC converters we conservatively assume operation at their worst-case efficiency of 50 \%.
To compare these numbers directly with our OPC UA engine, we subtract the power while serving the read or write node request from the idle power with network load. These estimated values representing only the CPU and DRAM power are shown in the second half of table \ref{tab:power-results-software}.
The software stack with the SoC running at \SI{600}{\mega\hertz} processes a read node (write node) request in \SI{723}{\micro\second} (\SI{700}{\micro\second}) and \SI{482}{\micro\second} (\SI{466}{\micro\second}) at \SI{900}{\mega\hertz}.
In summary we estimate that the SoC needs \SI{13.66}{\micro\joule} (\SI{13.98}{\micro\joule}) at \SI{600}{\mega\hertz} and \SI{16.48}{\micro\joule} (\SI{18.69}{\micro\joule}) at \SI{900}{\mega\hertz} for reading (writing) a node.

\begin{figure}
  \includegraphics[width=0.5\textwidth]{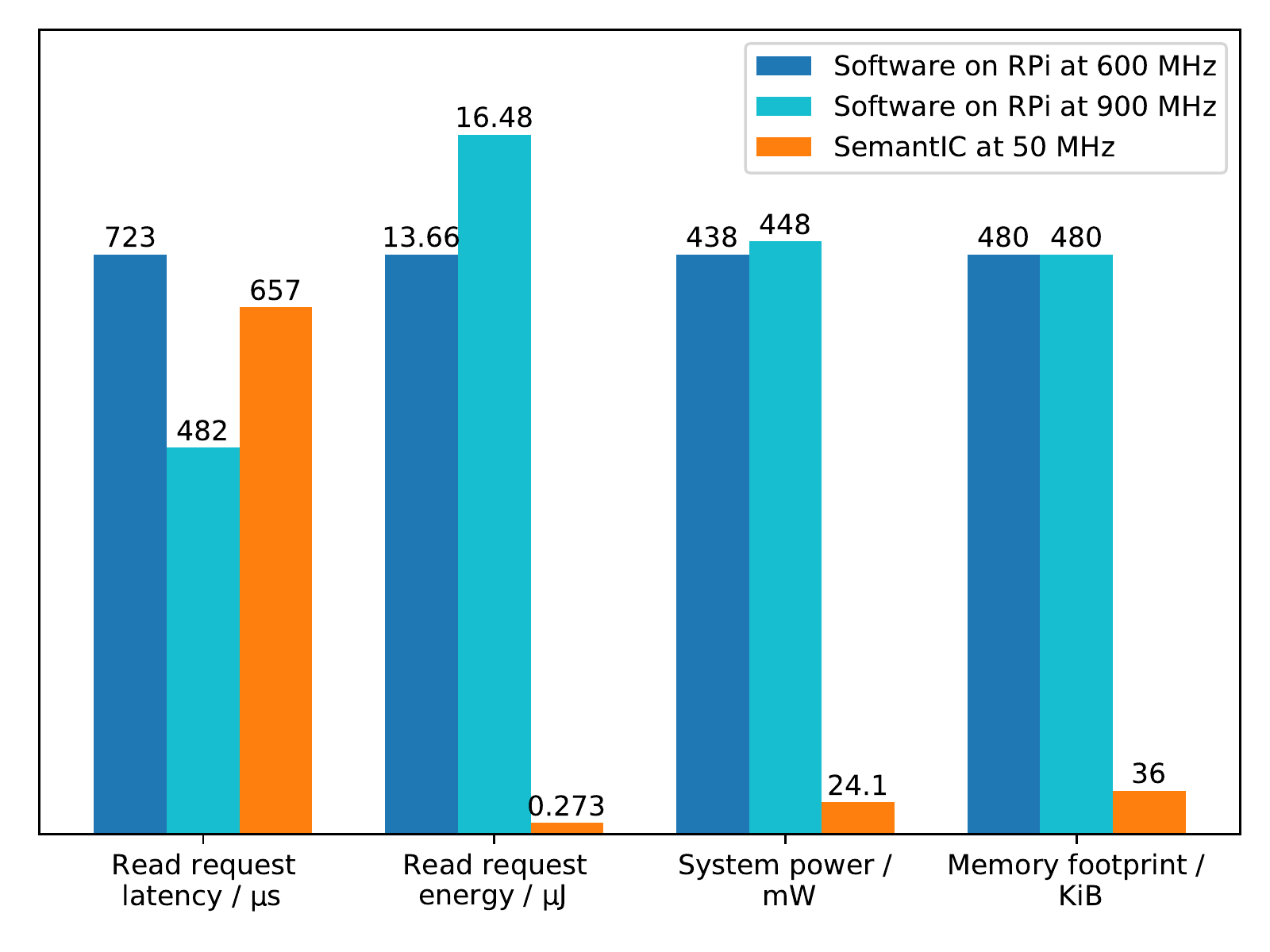}
  \caption{Comparison of the SemantIC test chip versus the open62541 software implementation on a Raspberry Pi 2B.}
  \label{fig:comparison}
\end{figure}

We summarize the main results as a comparison between our hardware implementation and the software in figure \ref{fig:comparison}. The OPC UA engine delivers comparable performance at \SI{50}{\mega\hertz} but is 50 times more energy efficient. On a system level, we estimate that the SoC (without the LAN9514 PHY) draws \SI{438.5}{\milli\watt} under load, 18 times more than our complete chip. The memory footprint is reduced by a factor of 13 to only 36 KiB, which is compatible with embedded platforms.

As an outlook for future improvements, we performed a trial synthesis of the OPC UA engine with cell libraries characterized for adaptive reverse body bias\footnote{For this test chip tapeout these libraries were not available yet.}. This methods is very effective at reducing leakage \cite{bib:Walter2020}, and would help scenarios with low duty cycles where the OPC UA engine remains mostly idle. Our preliminary results indicate that the OPC UA engine could be implemented at \SI{50}{\mega\hertz} and \SI{0.55}{\volt} with only \SI{50}{\micro\watt} leakage, a reduction by more than half.

\section{Conclusion}
We have presented the first ever ASIC implementation of an OPC UA server in a modern technology and provided detailed analysis of our test chip.
Measurements prove that our OPC UA hardware engine is superior to a popular software implementation in terms of power, energy, and memory consumption, while incurring only moderate cost in terms of silicon area.
There is further potential for future implementations to consider low-power optimization of the complete system, for example, by using adaptive reverse body bias.
We believe that dedicated hardware engines like ours are highly attractive and probably the only solution to bring OPC UA into every field device.

\label{sec:conclusion}
\section*{Acknowledgment}
We would like to thank Racyics GmbH for donating 22 nm foundation IP and for providing tapeout support.

%%%%%%%%%%%%%%%%%%%%%%%%%%%%%%%%%%%%%%%%%%%%%%%%%%%%%%%%%%%%%%%%%%%%%%%%%
%\nocite{*}

\bibliographystyle{IEEEtran}
\bibliography{paper_semantic}
%%%%%%%%%%%%%%%%%%%%%%%%%%%%%%%%%%%%%%%%%%%%%%%%%%%%%%%%%%%%%%%%%%%%%%%%%

\end{document}